\documentclass[conference,letterpaper]{IEEEtran}

\usepackage[utf8]{inputenc} 
\usepackage[T1]{fontenc}
\usepackage{url}
\usepackage{ifthen}
\usepackage{cite}
\usepackage[cmex10]{amsmath} 

\usepackage{amsfonts}
\usepackage{amssymb}
\usepackage{algorithmic}
\usepackage{mathrsfs}
\usepackage{array}
\usepackage{graphicx}
\usepackage{comment}
\usepackage{enumerate}
\usepackage{tikz}
\usepackage{enumitem}
\ifCLASSOPTIONcompsoc
  \usepackage[caption=false,font=normalsize,labelfont=sf,textfont=sf]{subfig}
\else
  \usepackage[caption=false,font=footnotesize]{subfig}
\fi

\usepackage{comment}
\usepackage{etoolbox}
\makeatletter
\patchcmd{\@makecaption}
  {\scshape}
  {}
  {}
  {}
\makeatother

\usepackage{bigdelim}
\usepackage{arydshln}

\newcommand{\BB}{\mathbb{B}}

\newcommand{\JJ}{\mathcal{J}}
\newcommand{\II}{\mathcal{I}}

\newcommand{\F}{\mathcal{F}}
\newcommand{\qed}{\hspace*{\fill}~{\rule{2mm}{2mm}}\par\endtrivlist\unskip}

\newcommand{\Z}{\mathbb{Z}}

\newcommand{\J}{\mathcal{J}}

\newcommand{\D}{\mathcal{D}}

\newcommand{\MC}{\mathcal{L}}

\newtheorem{lemma}{Lemma}
\newtheorem{theorem}{Theorem}

\newtheorem{problem}{Problem}

\newtheorem{defn}{Definition}

\DeclareMathAlphabet{\mathpzc}{OT1}{pzc}{m}{it}

\renewcommand{\vec}[1]{\underline{#1}}

\begin{document}
\title{Computing the Discrete Fourier Transform of signals with spectral frequency support}


\author{%
    \IEEEauthorblockN{P Charantej Reddy}
  \IEEEauthorblockA{IIT Hyderabad\\
                    ee18resch01010@iith.ac.in} \and
                    
  \IEEEauthorblockN{ V S S Prabhu Tej}
  \IEEEauthorblockA{IIT Hyderabad\\
                   ee19mtech11030@iith.ac.in} \and     
                    
  \IEEEauthorblockN{Aditya Siripuram}
  \IEEEauthorblockA{IIT Hyderabad\\
                    staditya@iith.ac.in}  \and
 
  \IEEEauthorblockN{Brad Osgood}
  \IEEEauthorblockA{Stanford University\\
                    osgood@stanford.edu}
}

\maketitle

\begin{abstract}
We consider the problem of finding the Discrete Fourier Transform (DFT) of $N-$ length signals with known frequency support of size $k$. When $N$ is a power of 2 and the frequency support is a spectral set, we provide an $O(k \log k)$ algorithm to compute the DFT. Our algorithm uses some recent characterizations of spectral sets and is a generalization of the standard radix-2 algorithm. 
\end{abstract}
\section{Introduction}\label{section:intro}
The Discrete Fourier Transform (DFT) represents a signal as a combination of complex exponentials (or frequencies). Analysis of signals using the DFT has been a popular tool in many areas of engineering and science. Given an $N-$ length signal $\vec{x}$, its DFT is another $N-$ length signal given by $\F_N \vec{x}$, where \[\F_N = \left(e^{-2\pi i mn/N}\right)_{m,n=0}^{N-1}\] is the $N\times N$ DFT matrix. While a naive multiplication of the DFT matrix $\F_N$ with the signal $\vec{x}$ would incur a computational complexity of $O(N^2)$, the Fast Fourier Transform (FFT) is a suite of algorithms that compute this multiplication (by exploiting the structure of the matrix $\F_N$) with $O(N\log N)$ computational complexity\footnote{By computational complexity we mean the number of (complex) additions and multiplications used by the algorithm. Also recall that we say the complexity is $O(g(N))$ to mean that for sufficiently large $N$ the number of operations required is bounded above by $c g(N)$ for some constant $c$.}. Perhaps the most well known among these algorithms is the radix-$2$ FFT \cite{1162257}, which assumes $N$ to be a power of $2$ and exploits the block structure of the matrix $\F_N$. Many other algorithms, including Cooley-Tukey FFT \cite{cooley1965algorithm}, Good Thomas FFT \cite{good1958interaction} and Rader's FFT \cite{rader1968discrete} (to name a few) developed in the literature can be used to compute the DFT in $O(N\log N)$, even when $N$ is non-prime power composite, or itself a prime. The importance of FFT makes it one of the most important algorithms developed in the last century\cite{10.2307/29775194,dongarra2000guest}.

Many applications of DFT rely on the crucial assumption that most of the frequency components (i.e., the entries of $\F_N\vec{x}$) are zero (or close to zero). This naturally leads to the question: if the frequency components are known to be zero at certain locations, can we do better than $O(N\log N)$? In this work, we attempt to address the following problem:

\begin{problem}
Given a space of $N-$length signals $\BB^\JJ$, such that for any signal $\vec{x} \in \BB^\JJ$ in the space, its DFT $\F_N\vec{x}$ is known to be nonzero only at locations $\JJ$:
\[
\F_N\vec{x}(j)=0 \quad \text{ for any }j\notin \JJ;
\]
What is the best achievable computational complexity for finding the DFT of signals in this space?
\end{problem}
Here we assume that $\JJ$ is known beforehand. We make some preliminary observations on this problem in Section \ref{section:pre_observations}: a naive computation of the $|\JJ|=k$ frequency coefficients incurs a complexity of $O(Nk)$, and with a simple argument, we can easily show that the DFT of such signals can be computed with $O(k^3)$ complexity, irrespective of the frequency support $\JJ$ (more on this in Section \ref{section:pre_observations}). However, it is also apparent with simple examples, that more structured $\JJ$ is, the lesser the complexity required to compute the DFT. We can take for instance, assuming $k$ divides $N$, the frequency support $\JJ$ to be periodic ( $\JJ=\{0,N/k, 2N/k, \ldots(k-1)N/k\}$), or a set of consecutive elements $\JJ =\{0,1,2,\ldots, k-1\}$: in both these cases we can argue (using basic properties of the DFT, again, in Section \ref{section:pre_observations}) that the DFT of signals with such frequency supports can be computed in $O(k \log k)$. We can then ask if there is any general structural property of the support set $\JJ$ that enables an $O(k \log k)$ computation of the DFT. 

Towards this end, we assume the set $\JJ$ is \emph{spectral}:
\begin{defn}
We say that a set $\JJ \subseteq \mathbb{Z}_N$ is spectral if there exists a set $\II \subseteq \mathbb{Z}_N$ of the same size as $\JJ$, such that the (square) Fourier submatrix of $\F_N$ with rows indexed by $\II$ and columns indexed by $\JJ$ is unitary\footnote{Here and in the rest of the document, whenever we say unitary, we mean unitary up to scaling.} (up to scaling). If $M$ is such a submatrix, it satisfies $M^*M = k I$ where $k=|\JJ|$ and $I$ is the $k \times k$ identity, and  $\mathbb{Z}_N$ is the set of integers modulo $N$ \footnote{ For a matrix $M$, $M^*$ denotes the conjugate transpose of $M$.}.
\end{defn}
 Indeed, the periodic and consecutive element sets mentioned earlier are examples of spectral sets. These sets are relevant in the context of Fuglede's conjecture \cite{fuglede1974commuting} in Fourier analysis, and this conjecture is as yet open for the discrete case for arbitrary $N$ \cite{dutkay2014some, siripuram2018lp}.

When $\JJ$ is spectral and $N$ is a power of $2$, we provide a (deterministic) algorithm to compute the DFT of signals in $\BB^\JJ$ that has complexity $O(k \log k)$. Our algorithm uses recent results on the structure of spectral sets \cite{fan2016compact,siripuram2019discrete,siripuram2020convolution} in terms of the binary expansion of the indices in $\JJ$. Our algorithm reads $k$ entries of the vector $\vec{x}$ at specific locations chosen according to the structure of the spectral set $\JJ$. The algorithm operates very similar to the radix-2 FFT algorithm (see Fig \ref{fig:block-diagram}): the crucial difference is that as opposed to the digit reversing permutation used by the radix-2 FFT, our algorithm reverses only a subset of digits. Since $\JJ = \mathbb{Z}_N$ is trivially a spectral set (Problem 1, in this case, reduces to finding the standard DFT), our algorithm can be seen as a generalization of the standard radix-2 FFT.

In Section \ref{section:motivation_SFFT}, we explain our motivations and try to place this result among other results in sparse FFT algorithms. In Section \ref{section:pre_observations}, we make some preliminary observations on Problem 1 and elaborate on some of the comments mentioned in the introduction. In Section \ref{section:main_result}, we present the main result (Theorem \ref{thm:block-structure}) that enables our algorithm to work, and finally in Section \ref{section:proofs}, we provide the proof for Theorem \ref{thm:block-structure}. Though we give our proof for the case when $N$ is a power of $2$ for ease of exposition; these techniques can be easily generalized to the case when $N$ is a power of any prime $p$.
\section{Motivations and connections to Sparse FFT literature}\label{section:motivation_SFFT}
The problem of efficiently finding the DFT is even more important considering the ever-increasing data sizes that emerging technologies generate and analyze. As such, much of the work on FFT in this century has focused on the sparse-FFT algorithms. Most of these algorithms assume the DFT $\F_N \vec{x}$ has only $k$ (typically $k << N$) non zero entries (but the locations of these $k$ non zero entries are unknown). In addition to computational complexity, algorithms in sparse DFT computation are also interested in minimizing the sample complexity, which is the number of the entries of $\vec{x}$ that the algorithm needs access to compute $\F_N \vec{x}$. What differentiates this area research from the allied areas of compressed sensing and sparse signal recovery \cite{2051150} is the emphasis on computational complexity, in addition to sample complexity. 

The best known sample complexity is $O(k \log N)$ \cite{14819549}, and the best known time complexity is $O(k \log N \log(N/k))$ \cite{nearlyoptimalsparse}; but it is not yet known if the same algorithm can achieve both of these. Some of these algorithms are probabilistic, which assume that the frequency support is more or less uniform and provide algorithms that work with constant or high probability for large $N$ \cite{gilbert2014recent, pawar-ramachandran, nearlyoptimalsparse}. 

More recently, there has been an increasing interest in developing algorithms for the case when the sparsity pattern is not completely arbitrary. Among these include results for block sparse signals \cite{3055462} that achieve a complexity of $O(k \text{ poly} (\log n))$. 

Along these lines, the problem that motivated us was the DFT computation for signals with partially known support, similar to such models in compressed sensing \cite{2051150}. An optimal algorithm in such a setting has to make use of the (partially known) support structure in some way. At the extreme, one can ask about the optimal complexity when the support is fully known: this leads us to Problem 1. To the best of our knowledge, Problem 1 has not been tackled from the computation perspective.

In this context, our algorithm is deterministic, assumes the frequency support is known and spectral and has a sample complexity $k$ and computational complexity $O(k \log k)$.

\section{Preliminary observations} \label{section:pre_observations}
Let us start with some simple observations on Problem 1. Given an $\vec{x}$, we will often refer to $\vec{x}$ as being in the time domain and $\F \vec{x}$ as being in the frequency domain. We will denote the DFT matrix with $\F_N$ or $\F$ when $N$ is apparent from the context. We also denote by $\vec{x}_\II$ the vector obtained by taking only the entries indexed $\II$ from $\vec{x}$, and by $A(\II,\JJ)$ the submatrix $A$ with rows indexed $\II$ and columns indexed $\JJ$. 

Now for $\vec{x}\in \mathbb{B}^\JJ$, we have 
\[
\vec{x} = \underbrace{\F^{-1}}_{\substack{\text{inverse }\\ \text{DFT matrix }}}\underbrace{\F \vec{x}}_{ \text{DFT of }\vec{x}},
\]
and since $\F \vec{x}$ (the signal in the frequency domain) is non-zero on locations in $\JJ$, only the columns of $\F^{-1}$ indexed by $\JJ$ (in other words, only the complex exponentials with frequencies from $\JJ$) play a role in the reconstruction of $\vec{x}$. Suppose we read only $k$ entries of the vector $\vec{x}$ corresponding to the locations $\II$ in the time domain, we get 
\[
\vec{x}_\II = \F^{-1}(\II,\JJ)(\F\vec{x})_\JJ.
\]
To find $(\F\vec{x})_\JJ$ (and hence $\F\vec{x}$), we could solve the above system of equations to get
\begin{equation}
\label{eq:recovery-from-time-samples}
(\F\vec{x})_\JJ = \left(\F^{-1}(\II,\JJ)\right)^{-1}\vec{x}_\II,
\end{equation}
provided we pick $\II$ in such a way that the resulting submatrix $\F^{-1}(\II,\JJ)$ is invertible. One easy way to obtain an invertible submatrix is to pick $\II$ to be $k$ consecutive elements of $\mathbb{Z}_N$, this ensures the resulting submatrix is Vandermonde and hence invertible \cite{donoho-stark:uncertainty, DS-1}. The solution to \eqref{eq:recovery-from-time-samples} involves inverting a $k \times k$ matrix, and hence has a complexity of $O(k^3)$ \cite{boyd2018introduction}. Note that this works irrespective of $\JJ$, and scales only with the size of $\JJ$ and not the dimension $N$. However, the $O(k^3)$ complexity is applicable only in the noiseless case, as the resulting Vandermonde matrix, though invertible, maybe poorly conditioned \cite{DS-1}.

As discussed in the introduction, we can consider specific frequency support sets $\JJ$: for e.g., suppose we assume $k$ divides $N$ and set $\JJ = \{0,1,2,\ldots, k-1\}$. We can then downsample the signal $\vec{x}$ in the time domain to take samples at $\II=\{0,k',2k',\ldots,(k-1)k'\}$ where $k'=N/k$. We know from elementary Fourier analysis that the DFT undergoes aliasing \cite{osgood2018lectures}:
\[
\F_k \vec{x}_\II (m) = \sum_{i}\F_N \vec{x}(m-ik), \quad m=0,1,2\ldots,k-1.
\]
but since the signal in the frequency domain is limited to $\{0,1,\ldots,k-1\}$ this aliasing does not lead to any overlaps. This reduces the complexity of finding $\vec{x}$ to the complexity of finding a $k-$ point DFT, resulting in $O(k \log k)$ complexity\footnote{A similar argument can be applied when $\JJ = \{0,k',2k',\ldots, (k-1)k'\}$ is a periodic set.}. Note that this can be seen directly from \eqref{eq:recovery-from-time-samples}: for the given $\II$ and $\JJ$, the resulting submatrix $\F^{-1}(\II, \JJ)$ has entries $\exp( 2\pi i mk' n/ N) = \exp(2\pi i mn/k)$ for $m,n = 0,1,\ldots, k-1$. Thus the submatrix $\left(\F^{-1}(\II,\JJ)\right)^{-1} = k'\F_k$ is the $k \times k$ DFT matrix and the $O(k \log k)$ complexity follows. 

We can then ask, what is the structure on $\JJ$ which enables $O(k \log k)$ computation of the DFT of signals in $\mathbb{B}^\JJ$? To start with, we can assume that it is possible to pick an $\II$ such that the submatrix $F^{-1}(\II, \JJ)$ is unitary, and this leads us directly to the definition of spectral sets. In this case, \eqref{eq:recovery-from-time-samples} reduces to 
\begin{equation}
    \label{eq:orthogonal-recovery-from-time-samples}
(\F\vec{x})_\JJ = k'\left(\F(\JJ,\II)\right)\vec{x}_\II.
\end{equation}
This assumption directly reduces the complexity to $O(k^2)$, however, unlike in the example above, the submatrix $\F(\JJ,\II)$ may not be a $k \times k$ DFT matrix.Take for example, $N=2^{10}$, $\JJ = \{1, 292, 641, 932\}$, and $\II = \{316, 384, 828, 896\}$, the resulting submatrix is\\

\resizebox{0.95\hsize}{!}{%
$\begin{pmatrix}
-0.36 - 0.93i &  -0.71 - 0.71i &   0.36 + 0.93i &  0.71 + 0.71i \\
0.77 - 0.63i &  -1.00 + 0.00i  &  0.77 - 0.63i &  -1.00 + 0.00i \\
0.36 + 0.93i &  -0.71 - 0.71i &  -0.36 - 0.93i  &  0.71 + 0.71i \\
-0.77 + 0.63i &  -1.00 + 0.00i &  -0.77 + 0.63i &  -1.00 + 0.00i 
\end{pmatrix}$}\\

Which can be checked to be unitary. However, this is not a $k\times k$ DFT matrix (nor can it be written as $D_1\F_kD_2$ for some diagonal matrices $D_1, D_2$). Thus, the $O(k \log k)$ complexity does not follow from the already known FFT algorithms.

Also, note that given a spectral set $\JJ$, there could be many possible time domain samples $\II$ that result in a unitary submatrix \cite{siripuram2019discrete}. For one specific choice of $\II$, we prove in Theorem \ref{thm:block-structure} that the resulting submatrix $\F(\JJ, \II)$ has a block structure (similar to $\F_k$) that enables $O(k \log k)$ computation (down from $O(k^2)$) in \eqref{eq:orthogonal-recovery-from-time-samples}. It is not yet clear to us if this property extends to arbitrary unitary submatrices of the DFT matrix.

\section{Main result}\label{section:main_result}
\begin{figure*}[ht]
\centering

\includegraphics[width=0.75\textwidth]{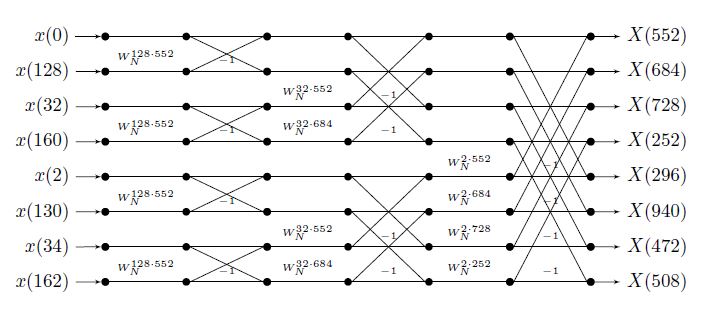}

\caption{DFT computation of $1024-$ length signal $x$ with spectral support set $\JJ = \{252, 296, 472, 508, 552, 684, 728, 940\}$ and sparsity $k = 8$}
\label{fig:block-diagram}
\end{figure*}

In this section, we present our main result. We will discuss about the structure of spectral sets, but first, we note that for $N$ is a power of 2, any spectral set has a size $k=2^r$ (see Lemma \ref{lem:spectral-structure}). The following theorem is the main result of this work:
\begin{theorem}
\label{thm:block-structure}
Suppose that $N$ is a power of $2$ and that $\JJ\subseteq \mathbb{Z}_N$ is a spectral set. Then under a suitable choice of indices $\II \subseteq \mathbb{Z}_N$, a suitable permutation of the rows and columns, the submatrix $\F(\JJ, \II)$ has the form 
\[
\begin{pmatrix}
I & D \\ I & -D
\end{pmatrix}\begin{pmatrix}
M & 0 \\ 0 & M
\end{pmatrix}
\]
where $D$ is a diagonal matrix; and $M$ is a unitary submatrix of $\F$ of size $k/2$.
\end{theorem}
 Recall that to solve \eqref{eq:orthogonal-recovery-from-time-samples}; we need to do the multiplication $\F(\JJ, \II)\vec{v}$. If $T(k)$ is the complexity to compute $\F(\JJ, \II)\vec{v}$, then from the structure of $\F(\JJ, \II)$ in Theorem \ref{thm:block-structure}, we have $T(k) = 2T(k/2) + 2k$. Since $k=2^r$ is a power of $2$, Theorem \ref{thm:block-structure} can be recursively applied and results in a complexity of $O(k \log k)$. Note that this is very similar to the block structure of the DFT matrix \cite{osgood2018lectures}: we may call the diagonal elements of $D$ as the twiddle factors (see Fig \ref{fig:block-diagram}).

To elaborate on the statement of Theorem \ref{thm:block-structure}, we need to explain the structure of spectral sets first. Consider writing, for each index in $\JJ$, the corresponding binary digits, arrayed in rows. The columns of such an arrangement represent the bits: starting with the least significant bit on the left, and each row of such an arrangement represents an index of $\JJ$ (see Fig \ref{fig:digit-table-example}).

Define the \emph{pivots} of such an arrangement to be the positions which contain the first (starting from the left) difference for some pair of rows. In Fig \ref{fig:digit-table-example}, for instance, indices $636$ and $545$ differ in the digit corresponding to $2^0$, whereas the indices $636$ and $1020$ first differ in the digit corresponding to $2^7$ (and are identical before that). Similarly, considering all the other differences, we see that $\{2^0, 2^7\}$ form the pivots. Finally, if we denote the pivots by $\MC$, we say a digit table is \emph{conforming} if $|\JJ| = 2^{|\MC|}$. 

\begin{figure}[htb] 
\begin{center}
\begin{tabular}{c|c|cccccc|c|cc}
\cline{2-2} \cline{9-9}
 & $2^0$ & $2^1$ & $2^2$ & $2^3$ & $2^4$ & $2^5$ & $2^6$ & $2^7$ & $2^8$ & $2^9$  \\
\hline
 636 & 0 & 0 & 1 & 1& 1 & 1 & 1 & 0 & 0& 1 \\
 545 & 1 & 0 & 0 & 0 & 0& 1 & 0 & 0 & 0& 1 \\
 1020 & 0 & 0 & 1 & 1 & 1 & 1 & 1 & 1 & 1& 1 \\
 161 & 1 & 0 & 0 & 0 & 0& 1 & 0 & 1 & 0& 0 \\
 \cline{2-2} \cline{9-9}
 \end{tabular}
 \end{center}
 \caption{An example digit table for $\JJ=\{161,545,636,1020\}$, $N=1024$. The rows of the digit-table represent the digits of an element of $\JJ$. Here $\MC = \{2^0,2^7\}$ (highlighted) represent the pivots. This is a conforming digit table, since $4=|\JJ|=2^{|\MC|}$. Note that the pivoted digits take all $4$ possible values $00, 10, 01$ and $11$.}
 \label{fig:digit-table-example}
\end{figure}

Suppose we consider the $|\MC|-$tuple representing the pivoted digits for each index. By definition of pivots, all these tuples must be distinct. Conformity enforces that these tuples must take all the possible $2^{|\MC|}$ values. For the example $\JJ$ in Fig \ref{fig:digit-table-example}, the pivot digits for $636,545,1020, 161$ are $00,10,01,11$ respectively. With these definitions in place, we have the following:
 \begin{lemma}
 \label{lem:spectral-structure}
 when $N$ is a power of $2$, the submatrix $\F^{-1}(\II,\JJ)$ is unitary iff for some set of pivots $\MC$
 \begin{enumerate}
     \item $\JJ$ corresponds to a conforming digit table with pivots $\MC$,
     \item $\II$ corresponds to a conforming digit table with pivots $N/2\MC$.
 \end{enumerate}
 \end{lemma}
Lemma \ref{lem:spectral-structure} is a direct consequence of earlier works \cite{siripuram2019discrete, siripuram2020convolution}; we discuss more about this in Section \ref{section:proofs}.

Suppose we start with a $\JJ$ that is spectral, construct its digit table, and read off the pivot columns $\MC$. By Lemma \ref{lem:spectral-structure}, this must be a conforming digit table, and we must have $|\JJ| = 2^{|\MC|}$. Now we take $\MC' = N/2\MC$ and construct a digit table by setting the pivoted digits to take all possible $|\MC'|-$tuples and all non pivoted digits to zero. For e.g., consider the set $\JJ$ in Fig \ref{fig:digit-table-example}: the pivots are $\MC=\{2^0,2^7\}$; so we set $\MC' = N/2\MC = \{2^2, 2^9\}$, and construct the digit-table, leading to Fig \ref{fig:I-digit-table-example}. This construction ensures the resulting digit-table is conforming. We then take the indices corresponding to this digit table as the time domain sampling locations $\II$. This is the choice of indices $\II$ referred to in the statement of Theorem \ref{thm:block-structure}.
\begin{figure}[htb] 
\begin{center}
\begin{tabular}{ccc|c|cccccc|c|}
\cline{4-4} \cline{11-11}
 & $2^0$ & $2^1$ & $2^2$ & $2^3$ & $2^4$ & $2^5$ & $2^6$ & $2^7$ & $2^8$ & $2^9$  \\
\hline
 0 & 0 & 0 & 0 & 0& 0 & 0 & 0 & 0 & 0& 0 \\
 512 & 0 & 0 & 0 & 0 & 0& 0 & 0 & 0 & 0& 1 \\
 4 & 0 & 0 & 1 & 0 & 0 & 0 & 0 & 0 & 0& 0 \\
 516 & 0 & 0 & 1 & 0 & 0& 0 & 0 & 0 & 0& 1 \\
 \cline{4-4} \cline{11-11}
 \end{tabular}
 \end{center}
 \caption{Construction of the time domain samples $\II$ for the example in Fig \ref{fig:digit-table-example}. The non pivot digits are set to zero, and the pivot digits take all possible values.}
 \label{fig:I-digit-table-example}
\end{figure}

For the structure in Theorem \ref{thm:block-structure} to be realized, we also need an appropriate sorting of the indices in $\II$ and $\JJ$. The sorting on $\II$ and $\JJ$ is related to the pivot digits:
\begin{enumerate}
    \item The indices of $\II$ (i.e., the columns of the submatrix $\F(\JJ, \II)$) are sorted lexicographically based on the pivot digits, starting from the left to the right. For example, with two pivots (as in Fig \ref{fig:I-digit-table-example}), the index with pivot digits $00$ comes first, followed by the indices with pivot digits $01$, $10$, and $11$ respectively.
    \item The indices of $\JJ$ (i.e., the rows of the submatrix $\F(\JJ, \II)$) are sorted lexicographically based on the pivot digits, starting from the right to the left. For example, with two pivots (as in Fig \ref{fig:digit-table-example}), the index with pivot digits $00$ comes first, followed by the indices with pivot digits $10$, $01$ and $11$ respectively.  The non pivot digits are ignored for the purpose of sorting.
\end{enumerate}
For e.g., if $\JJ =\II = \mathbb{Z}_N$, then all the digits are pivots. In this case, the rows $\JJ$ are sorted in the natural order, whereas the columns $\II$ are sorted in the bit-reversed order, as in the standard radix-2 FFT (\cite{oppenheim1999discrete, osgood2018lectures}). Our algorithm here can be seen as a generalization of the same.

With the specific choice of $\II$, and the sorting on the rows and columns of the submatrix $\F(\JJ, \II)$ mentioned above, the structure in Theorem \ref{thm:block-structure} applies. We defer the proof to Section \ref{section:proofs}.


\section{Proofs}\label{section:proofs}
In this section, we will prove the results presented earlier. For this, we will find it convenient to make some definitions first. Suppose $1_\II, 1_\JJ \in \mathbb{C}^N$ represents the indicators of the sets $\II$ and $\JJ$. For $i \in \mathbb{Z}_N,$ we denote by $(i,N)$ the greatest common divisor (gcd) of $i$ and $N$. Define
\begin{equation}
\label{eq:def-idempotents}
    h_\II = \F^{-1}1_\II, \quad h_\JJ=\F^{-1}1_\JJ.
\end{equation}
These are convolution idempotents (i.e., they satisfy $h*h=h$ where $*$ is the discrete circular convolution). These idempotents arise naturally when taking inner products of any two columns of $\F(\JJ, \II)$, as in the lemma below.

\begin{lemma}
\label{lem:unitary-idempotent}
The submatrix $\F(\J,\II)$ is unitary iff $h_\JJ(i_1-i_2) = 0$ for any $i_1, i_2\in \II$ ($i_1\neq i_2$),
\end{lemma}
\begin{IEEEproof}
The inner product of any two columns indexed $n_1$ and $n_2$ is given by \(\sum_{m \in \JJ}\exp(2 \pi i m(n_1-n_2)/N = Nh_\JJ(n_1-n_2).\)
\end{IEEEproof}
So we want to construct $\II$ in such a way that the differences of any two elements in $\II$ fit in the zero set of $h_\JJ$. However, the zero set of $h_\JJ$ correspond to the roots of a polynomial with integer coefficients; and as such, they have a lot of structure. In particular, we have $h_\JJ(n)=0$ iff $h_\JJ(ns)=0$ for any $s$ coprime to $N$. This allows us to write the zero set of $h_\JJ$ as \(\{i \in \Z_N  \colon (i,N) \in \D_\JJ\}\), where $\D_\JJ$ is some set of divisors of $N$. Thus, the zero set contains all the indices in $\Z_N$ whose gcd with $N$ is in $\D_\JJ$. The proof is elementary: we direct the interested reader to the  references (\cite[Theorem~2.1]{malikiosis2016fuglede},\cite{siripuram2018lp}, \cite{siripuram2020convolution, reddy2020some}) for the proof and details. We refer to the set $\D_\JJ$ often as zero-set divisors of $h_\JJ$.

The involvement of gcd is why digit tables are very convenient to represent spectral sets. For any two indices $i_1, i_2$, we see that the gcd $(i_1-i_2, N)$ relates to the first non zero difference in the digits of $i_1$ and $i_2$. For any set $\II$, the set $\{(i_1-i_2,N):i_1, i_2 \in \II, i_1 \neq i_2\}$ is simply the set of pivots in the digit table for $\II$. This is the crucial observation we use next.

\subsection{Proof of Lemma \ref{lem:spectral-structure}}
Starting with $\JJ$, as in the preceding discussion, let $D_\JJ$ be the zero-set divisors of $h_\JJ$. Then $\F(\JJ,\II)$ is unitary, iff $(i_1-i_2,N) \in \D_\JJ$ for any $i_1, i_2\in \II$ (from Lemma \ref{lem:unitary-idempotent}). From the observations made previously, this means the digit table for $\II$ must have some subset of $\D_\JJ$ as pivots. The largest possible size of $\II$ (by the definition of pivots) is $2^{|\D_\JJ|}$: so we have $|\II| \leq 2^{|\D_\JJ|}$.

The rest of the proof of Lemma \ref{lem:spectral-structure} relies on the results on idempotents from \cite[Theorem 1]{siripuram2020convolution}: an idempotent $h_\JJ$ has zero set divisors $D_\JJ$ iff it is a concatenation of conforming digit tables with pivots $\MC^*=N/2\D_\JJ$. Since there must be at least one table in this concatenation, the size of $\JJ$ is at least $2^{|\MC^*|}$: so we have $|\JJ| \geq 2^{|\MC^*|} = 2^{|\D_\JJ|}$. However, $|\II| = |\JJ|$ and so combining the inequalities obtained till now gives $|\II| = |\JJ| = 2^{|\D_\JJ|}$. Consequently, the digit tables for both $\II$ and $\JJ$ are conforming: with pivots $\D_\JJ$ for $\II$ and pivots $N/2\D_\JJ$ for $\JJ$.
\qed
\subsection{Proof of Theorem \ref{thm:block-structure}}
\begin{figure*}[ht]
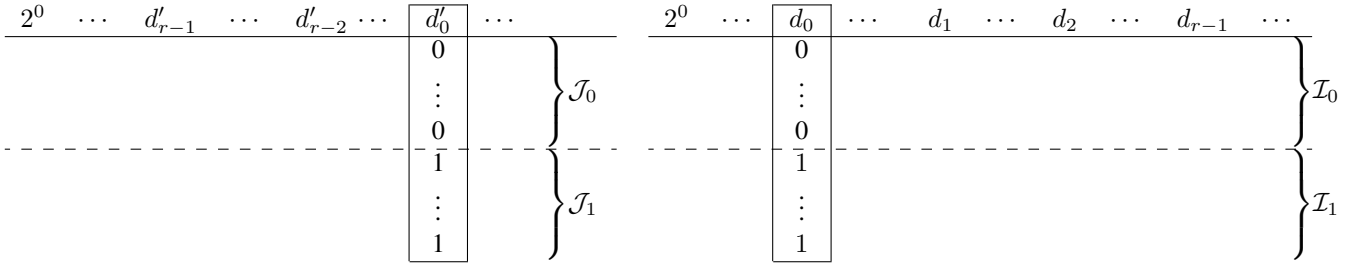

\centering
\begin{tabular}{cc}
			\begin{tabular}{ccccc|c|ccc}
			\cline{6-6}
			$2^0$  & $\cdots$ & $d'_{r-1}$  & $\cdots$ & $d'_{r-2}$  $\cdots$& $d'_{0}$& $\cdots$\\
			\cline{1-9}
			 & &  & &  &  0& & &  \hspace{-.2in}\rdelim\}{3.5}{*}[$\JJ_0$]\\ 
			 & &   & & &  \vdots&  &  &\\
			&  &  & &  &  0& &  &\\
			\cdashline{1-9}
			&  & &  & &  1& & & \hspace{-.2in}\rdelim\}{3.5}{*}[$\JJ_1$]\\
		    &  &   & &  & \vdots& &  &\\
			&  &  &  &   & 1& &  &\\
			\cline{6-6}
		\end{tabular}&
	\begin{tabular}{cc|c|cccccccc}
			\cline{3-3}
			$2^0$  & $\cdots$ & $d_0$  & $\cdots$ \, & $d_1$ & $\cdots$ & $d_2$ & $\cdots$& $d_{r-1}$& $\cdots$\\
			\cline{1-10}
			 & & 0 & &  & & & & & &  \hspace{-.2in}\rdelim\}{3.5}{*}[$\II_0$]\\ 
			 & & \vdots  & & & & & & & &\\
			&  & 0 & &  & & & & & &\\
			\cdashline{1-10}
			&  & $1$ &  & & & & & & & \hspace{-.2in}\rdelim\}{3.5}{*}[$\II_1$]\\
		    &  & \vdots  & & &  & & & &  &\\
			&  & $1$ &  & & &  & & & &\\
			\cline{3-3}
		\end{tabular}

\end{tabular}		
\caption{On the left: A general representation of confirming digit table (sorted) of frequency support set $\JJ$, here $d'_{0}$ is the largest pivot. On the right: Corresponding confirming digit table of time domain support set $\II$ constructed and sorted as explained in the Section \ref{section:main_result}. Here the smallest pivot is $d_{0} =N/2d_0'$.}
		\label{fig:digit-table-ordering}
		
\end{figure*}

This proof relies heavily on the ordering of $\II$ and $\JJ$ introduced in Section \ref{section:main_result}.
Some notation before we proceed: suppose $k=2^r$, and that the pivots for $\II$, from left to right, are $d_0, d_1, \ldots, d_{r-1}$. Then the pivots for $\JJ$, from left to right, are $d'_{r-1}, d'_{r-2}, \ldots, d'_0 $, where $d'=N/2d$ (this follows from Lemma \ref{lem:spectral-structure}). The ordering of $\II$ and $\JJ$ splits them naturally into smaller sets as in Figure \ref{fig:digit-table-ordering}: we have $\II = \II_0 \cup \II_1$ where $\II_0$ contains all indices with the leftmost pivot digit zero, and $\II_1$ contains all indices with the leftmost pivot digit one. A similar split $\JJ = \JJ_0 \cup \JJ_1$ occurs for the row indices $\JJ$, with the split based on the rightmost pivot digit.

This results in 
\begin{equation}
\label{eq:block-split}
\F(\JJ, \II) = \begin{pmatrix}
\F(\JJ_0, \II_0) & \F(\JJ_0, \II_1) \\ \F(\JJ_1, \II_0) & \F(\JJ_1, \II_1) 
\end{pmatrix}
\end{equation}

First, we note that both $\II_0$ and $\II_1$ have pivots $d_1, d_2, \ldots, d_{r-1}$; and both $\JJ_0$ and $\JJ_1$ have pivots $d'_{r-1}, d'_{r-2}, \ldots, d'_1$.  Since all these four sets have sizes $k/2 = 2^{|\MC \setminus \{d_0\}|}$, it follows that all these four sets correspond to conforming digit tables. From Lemma \ref{lem:spectral-structure}, it follows that all the four submatrices in \eqref{eq:block-split} are unitary.

We also make the following observations:
\begin{enumerate}[label=(\alph*)]
    \item All the entries of $\II$ are multiples of $2^{d_0}$. The entries of $\II_0$ are even multiples of $2^{d_0}$, whereas the entries of $\II_1$ are odd multiples of $2^{d_0}$.
    \item The digit tables for $\II_0$ and $\II_1$ are identical except for the leftmost pivot ($d_0$): thus $\II_1 = 2^{d_0} + \II_0$.
    \item The digit tables for $\JJ_0$ and $\JJ_1$ are identical up to the last pivot digit: the entries in $\JJ_0$ has the last pivot digit as $0$, whereas the entries in $\JJ_1$ have the last pivot digit as $1$.
    
    To see this, note that due to the proposed sorting, the $i^{\textsf{th}}$ in $\JJ_0$ and the $i^{\textsf{th}}$ index in $\JJ_1$ will have the same digits in all pivots except the last ($d'_0$). From the definition of pivots, this forces all the digits (including the non pivot digits) before $d'_0$ to be identical.
\end{enumerate}
Now to prove that $\F(\JJ,\II)$ has the structure of Theorem \ref{thm:block-structure}, we show the following
\begin{enumerate}
    \item $\F(\JJ_0, \II_1) = D \F(\JJ_0,\II_0)$\\
    From observation (b) above, we can write
    \begin{align*}
    \F(\JJ_0, \II_1) &= \left(e^{2\pi i mn/N} \right) \quad m \in \JJ_0, n \in \II_1 \\&= \left(e^{2\pi i m(n+2^{d_0})/N} \right) \quad {m \in \JJ_0, n \in \II_0} \\ &=D \F(\JJ_0,\II_0),
    \end{align*}
    where $D$ is a diagonal matrix with entries $\exp(2 \pi i m2^{d_0}/N)$, for $m \in \JJ_0$.
     \item $F(\JJ_1, \II_0) =  \F(\JJ_0,\II_0)$ \\
     We simply take the ratio of corresponding entries of these matrices and show that the ratio is $1$. This ratio is of the form
     \[
     e^{2\pi i m_{0j}n/N} / e^{2\pi i m_{1j}n/N}=     e^{2\pi i n(m_{0j} -m_{1j})/ N},
     \]
     where $n \in \II_0$, and $m_{00},m_{01} ,m_{02}, \ldots,$ are (in order) the entries of $\JJ_0$; similarly $m_{10},m_{11} ,m_{12}, \ldots,$ are (in order) the entries of $\JJ_1$. We note from observation (c) above that $m_{0j}-m_{1j}=\alpha 2^{d'_0}$, where $\alpha$ is odd. Further, from observation (a) above, the index $n$ is an even multiple of $2^{d_0}$, resulting in the ratio being
     \[
     e^{2\pi i \text{(even)}2^{d_0}\alpha 2^{d'_0}/ N} = 1.
     \]
      \item $\F(\JJ_1, \II_1) = - \F(\JJ_0, \II_1)$ \\
      This is similar to 2) above. We take the ratio of corresponding entries we get
      \[
     e^{2\pi i m_{0j}n/N} / e^{2\pi i m_{1j}n/N}=     e^{2\pi i n(m_{0j} -m_{1j})/ N},
     \]
     where $n \in \II_1$, and $m_{ij}$ are as before. From observation (a) above, $n$ is an odd multiple of $2^{d_0}$, so the ratio becomes
     \[
     e^{2\pi i \text{(odd)}2^{d_0}\alpha 2^{d'_0}/ N} = -1.
     \]
    \qed
\end{enumerate}





\end{document}